\documentclass[letterpaper,twocolumn,fleqn]{article} 

\usepackage{ist}
\usepackage{amsmath}
\usepackage{algorithm}
\usepackage{algorithmic}
\usepackage{tabularx}
\usepackage{multirow}
\usepackage{url}
\usepackage{flushend}

\title{A Fingerprint for Large Language Models}

\author{Zhiguang Yang, Shanghai University, Shanghai 200444, China; Email: yangzg@shu.edu.cn\\
Hanzhou Wu, Shanghai University, Shanghai 200444, China; Email: h.wu.phd@ieee.org (Corresponding author)}

\date{} 

\begin{document} 

\maketitle 

\thispagestyle{empty} 


\begin{abstract}
Recent advances confirm that large language models (LLMs) can achieve state-of-the-art performance across various tasks. However, due to the resource-intensive nature of training LLMs from scratch, it is urgent and crucial to protect the intellectual property of LLMs against infringement. This has motivated the authors in this paper to propose a novel black-box fingerprinting technique for LLMs. We firstly demonstrate that the outputs of LLMs span a unique vector space associated with each model. We model the problem of fingerprint authentication as the task of evaluating the similarity between the space of the victim model and the space of the suspect model. To tackle with this problem, we introduce two solutions: the first determines whether suspect outputs lie within the victim's subspace, enabling fast infringement detection; the second reconstructs a joint subspace to detect models modified via parameter-efficient fine-tuning (PEFT). Experiments indicate that the proposed method achieves superior performance in fingerprint verification and robustness against the PEFT attacks. This work reveals inherent characteristics of LLMs and provides a promising solution for protecting LLMs, ensuring efficiency, generality and practicality.
\end{abstract}

\section{Introduction}
Large language models (LLMs) have become a foundational component of contemporary artificial intelligence, exhibiting exceptional performance across a broad spectrum of natural language processing tasks owing to their capacity to generate human-like text and comprehend complex semantics. Despite the considerable data volume and computational power required for their training, numerous developers continue to advance the field by releasing their models as open source. Research groups and organizations responsible for prominent models, such as LLaMA, Gemma, and Mistral \cite{Touvron:arXiv:2023, Touvron:arXiv:2023:2, Mesnard:arXiv:2024, Jiang:arXiv:2023}, have made their well-trained LLMs publicly accessible, thereby fostering a dynamic and collaborative research ecosystem. This rapidly evolving landscape is largely sustained by its open-source philosophy. Nevertheless, the same openness that accelerates innovation also renders these models vulnerable to misuse, e.g., unauthorized fine-tuning with other pre-trained models without proper attribution, or misappropriation through false claims of ownership. Consequently, safeguarding the intellectual property rights of LLMs is imperative not only for preserving their commercial value but also for promoting the sustainable and ethical development of the community.

Numerous studies have employed digital watermarking and fingerprinting methods to protect deep neural networks (DNNs). For example, Uchida \emph{et al}. \cite{Uchida:ICMR:2017} introduce a regularization term to constrain the network weights for embedding watermarks. Subsequently, researchers have proposed various methodologies such as  \cite{Wang:EI:2020, Rouhani:ASPLOS:2019, Fernandez:ICASSP:2024} to embed watermarks into the model parameters in white-box scenarios, where the extractor can access the entirety of model parameters. However, the challenges in accessing all the model parameters in various scenarios has led to a focus on black-box techniques for model watermarking. To this purpose, many watermarking methods utilize backdoor techniques, constructing specific input-output mappings and observing the output of the DNN model for verification such as \cite{Adi:USENIX:2018, Zhao:ISDFS:2021, Liu:TDSC:2024}. Generative models, particularly image processing models, often produce contents with high entropy and sufficient information capacity to accommodate additional watermark information, which remains highly imperceptible. Wu \emph{et al}. \cite{Wu:TCSVT:2021} introduce a new framework to make the output of a model contain a certain watermark. By extracting the watermark from any watermarked output, one can identify the ownership of the corresponding DNN model. Lukas \emph{et al}. \cite{Lukas:USENIX:2023} propose embedding watermarks by fine-tuning the image generator, ensuring that all images produced are watermarked. Fernandez \emph{et al}. \cite{Fernandez:ICCV:2023} extend it to the diffusion model. Embedding watermarks through backdoor and fine-tuning techniques compromises the primary functionality of the model to a certain extent and requires significant computational resources. Song \emph{et al}. \cite{Song:CVPR:2024} distinguish different generative models based on their artifacts and fingerprints, which can help alleviate this problem.

The emergence of superior reasoning capabilities in LLMs which require significant computational overhead, has posed new challenges for their intellectual property protection (IPP). Xu \emph{et al}. \cite{Xu:NAACL:2024} specify a confidential private key and embed it as an instructional backdoor, serving as a fingerprint. In Ref. \cite{Zeng:NIPS:2024}, Zeng \emph{et al}. utilize the internal parameters in Transformer \cite{Vaswani:Transformer:2017} as a fingerprint to identify the LLMs. Existing methods typically require white-box access or fine-tuning to verify the copyright information. In contrast, the proposed fingerprinting method can be implemented in a black-box scenario without any fine-tuning.

We propose a novel fingerprinting approach for LLMs that enables LLM authentication through analysis of their output characteristics. LLMs generate coherent and contextually appropriate text by sampling from logits, which inherently encode rich model-specific information. In black-box scenarios, LLM providers often expose full or partial logit vectors via APIs, allowing users to employ different sampling strategies to produce realistic content. Carlini et al. \cite{Carlini:ICML:2024} have demonstrated that it is possible to recover parts of a model solely through such API access. Building on this insight, we introduce an LLM fingerprinting framework that identifies distinctive properties of each model by analyzing its logit outputs from a new perspective.

Prior studies have demonstrated that the outputs of LLMs lie within a linear subspace determined by their parameters (see \cite{Yang:ICLR:2018, Finlayson:ICLR:2024, Finlayson:COLM:2024}). In our approach, we preserve the parameters of the victim model - specifically those of its final linear layer. By querying the suspect model and collecting its outputs, we leverage the retained parameters to characterize the model's unique attribution. We formalize ownership authentication as the process of measuring the similarity between the vector space of the victim model and the output space of the suspect model.

We first introduce a method for rapidly determining whether a given output originates from the victim model by evaluating its compatibility with the vector space defined by the retained parameters. To handle scenarios involving Parameter-Efficient Fine-Tuning (PEFT) attacks, we further design an alignment verification approach that assesses whether the suspect model was derived from the victim model through PEFT by comparing their representational similarities. Moreover, in cases where only partial logits are accessible via an API, we demonstrate that complete logits can be reconstructed to facilitate ownership verification. Our framework enables model ownership authentication in black-box settings solely through API access, without relying on any specific architectural assumptions about the underlying LLM. This ensures broad applicability and offers a practical and effective solution for the copyright protection of LLMs. Experimental results show that our method achieves superior verification accuracy and robustness against PEFT attacks, without impairing the functional performance of the models.

In summary, the main contributions of this work include:
\begin{itemize}
	\item We analyze the characteristics of LLMs from a novel perspective and demonstrate that their outputs can serve as unique fingerprints for ownership verification.
	\item We propose two complementary methods for LLM ownership verification: one enables rapid identification of whether an output originates from the victim model, while the other determines whether a suspect model has been derived from the victim model through a PEFT attack.
	\item We introduce a technique to reconstruct complete fingerprint information from partial logits obtained via API access, enabling ownership verification in black-box scenarios.
\end{itemize}

The rest structure of this paper is organized as follows. First of all, we provide an overview of model fingerprinting, PEFT, and the assumed threat model. Then, we introduce the proposed LLM fingerprinting technique, followed by two fingerprint verification approaches. Thereafter, experimental results and analysis are provided. Finally, we conclude this paper and provide discussion.

\section{Preliminaries}
In this section, we provide a brief overview of model fingerprinting, PEFT, and the assumed threat model so that the proposed work can be better described.

\subsection{Model Fingerprinting}
Model fingerprinting is a technique used to uniquely identify and authenticate DNN models by analyzing their inherent characteristics. Every DNN model, even those with the identical architecture, develops distinctive patterns in its parameters and outputs due to differences in training data, initialization, and optimization. Through analyzing these patterns such as logits, embeddings, or behavioral responses to specific inputs, model fingerprinting can generate a ``signature'' that distinguishes one model from another. This technique enables applications such as ownership verification, detection of unauthorized copies or fine-tuned derivatives, and security auditing, and it can often be applied even in black-box scenarios where only model outputs are accessible. Through these methods, model fingerprinting provides a robust and practical way to trace, verify, and protect DNN models. In this research, we utilize the outputs of LLMs as fingerprints for verification.

\subsection{Parameter-Efficient Fine-Tuning (PEFT)}
Training or fine-tuning DNNs from pre-trained models requires substantial computational resources, especially for LLMs that demand extensive GPU memory. Recent research has therefore focused on parameter-efficient fine-tuning (PEFT) methods, which enable effective optimization while updating only a small subset of model parameters \cite{Houlsby:ICML:2019, Hu:ICLR:2022, Dettmers:NIPS:2023}. In other words, PEFT is a set of techniques designed to adapt large pre-trained models to specific tasks without updating all of the model's parameters. 

Instead of fine-tuning the entire model, which can be computationally expensive and memory-intensive, PEFT typically modifies only a small subset of parameters or adds lightweight modules, while keeping the majority of the model frozen. This makes fine-tuning faster, cheaper, and more storage-efficient.

Low-Rank Adaptation (LoRA) \cite{Hu:ICLR:2022} has become the de facto method for PEFT, serving as the foundation for many other methods such as \cite{Dettmers:NIPS:2023, Meng:NIPS:2024}. In LoRA, the weight matrix $\textbf{W}_\text{O}$ is updated by the formula $\textbf{W}_\text{N} = \textbf{W}_\text{O} + \Delta \textbf{W} = \textbf{W}_\text{O} + \textbf{A}\textbf{B}$. During training, $\textbf{W}_\text{O}$ remains fixed, while the two matrices $\textbf{A}$ and $\textbf{B}$ encompass the least trainable parameters. In this study, we will use LoRA to mimic the PEFT attack.

\subsection{Threat Model}
Our threat model involves a defender, known as the model provider, and an adversary who controls a malicious user. The adversary's goal is to steal the model and falsely claim ownership. We assume that the adversary may employ PEFT attacks such as LoRA \cite{Hu:ICLR:2022}, to evade detection. The defender, who retains access to their own model's parameters, including the final linear layer as a fingerprint, aims to verify ownership by querying the suspect model through API access.

\begin{figure}[!t]
	\centering
	\includegraphics[width=\linewidth]{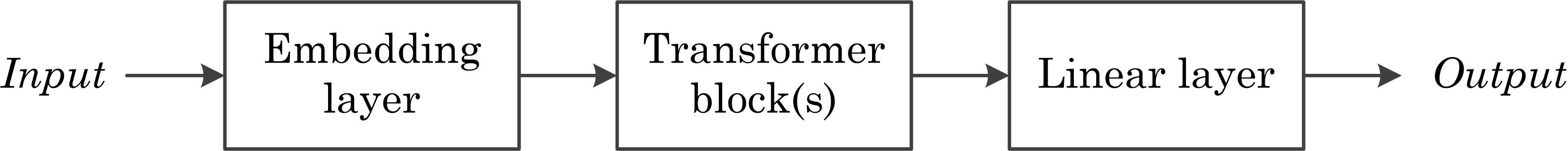}
	\caption{The pipeline of LLMs based on the Transformer architecture.}\label{pipeline}
\end{figure}

\section{LLM Fingerprinting}
In this section, we show that the logits outputs from LLMs span a vector space, which can be utilized for fingerprinting LLM.

\subsection{LLM Outputs Span A Vector Space}
The Transformer architecture has already become the base of numerous models due to its exceptional performance across a variety of tasks. LLMs are typically Transformer-based models, with their pipeline presented in Figure \ref{pipeline}. Specifically, the input text is tokenized and converted into word embeddings, with the embedding layer represented as a matrix of size $|V| \times h$, where $|V|$ is the vocabulary size and $h$ is the hidden size. The vocabulary size is significantly larger than the hidden size, i.e., $|V| >> h$. The embeddings are processed by the Transformer block with multiple layers, each containing a multi-head self-attention mechanism and a feed-forward layer, to calculate the intrinsic representation of the input and produce the intermediate representation $\textbf{z}$. The last linear layer maps $\textbf{z}$ to logits $\textbf{s} \in \textbf{R}^{|V|}$. The output of LLM is generated by sampling from the logits $\textbf{s}$. For clarity, this analysis excludes Layer Normalization \cite{Ba:arXiv:2016} and RMS Normalization \cite{Zhang:NIPS:2019}, as they primarily introduce additional multiplication terms, which do not affect the conclusions of this paper. 

It is worth noting that the logits $\textbf{s} = \textbf{W}\textbf{z}$, where $\textbf{W} \in \mathbf{R}^{|V| \times h}$ is the weight matrix of the last linear layer and the rank of $\textbf{W}$ is at most $h$. Every output produced by $\textbf{W}$ is corresponding to a vector $\textbf{s}$ that will always lie within a subspace of $\mathbf{R}^{|V|}$, spanned by the columns of $\textbf{W}$, which is at most $h$-dimensional. All possible outputs of the LLM logits will span a vector space $L$ isomorphic to the space spanned by the columns of $\textbf{W}$, since they have the same dimensionality. Consequently, each LLM is associated with a distinctive vector space $L$. This uniqueness arises because the vector space $\mathbf{R}^{|V|}$ encompasses an exceedingly large number of potential subspaces for any subspace of dimension $h$. For example, in the case of Gemma \cite{Mesnard:arXiv:2024}, with $h = 2,048$ and $|V| = 256,000$, the number of subspaces is extremely large. Due to variations in training initialization, datasets, configurations and hardware, it is impossible for two different LLMs to cover the same vector space. This property allows us to use it as a fingerprint for LLMs and identify their ownership. The model providers only need to retain the parameters of the last linear layer in the model. Accessing the API of the suspect model enables them to retrieve the logits, thereby verifying the ownership.

\subsection{Vector Space Reconstruction via API}
We already demonstrate that the logits outputs from LLMs span a vector space denoted by $L$, which will be utilized as the LLM fingerprint. Obtaining the full logits of a model is not always feasible, as attackers aim to disclose minimal information to evade detection by the victim. In this subsection, we consider practical scenarios in which our approach reconstructs the vector space for fingerprint verification and only relies on API access, enabling the retrieval of complete vocabulary probabilities, top-$k$ probabilities, or the top-$1$ probability.

\subsubsection{Complete Probabilities of the Vocabulary} 
The API provides the complete probabilities $\textbf{p} = \text{softmax}(\textbf{s})$ over the vocabulary, all components of which are non-negative and have a sum equal to 1. This indicates that $\textbf{p}$ is a point in the simplex $\Delta^{|V|-1}$, which lies within a $|V|-1$ dimensional subspace of $\mathbf{R}^{|V|}$. Additionally, $\textbf{p}$ is also constrained by $\textbf{s}$.

Due to normalization, the softmax function does not have a well-defined inverse transformation. However, if we omit it temporarily, and use the $\texttt{CLR}$ (centered log-ratio) transformation, we can obtain $\textbf{s}^\star$ that differs from $\textbf{s}$ by a constant bias. This will introduce a one-dimensional deviation from the perspective of spatial dimensions. We can manually do one-dimensional correction and will not affect the results, as $h$ and $|V|$ are both significantly greater than one. Therefore, we can directly reconstruct the vector space $L^\star$ corresponding to $\textbf{s}^\star$, which is determined by
\begin{equation}
\textbf{s}^\star = \texttt{CLR}(\textbf{p}) = \log\left(\frac{\textbf{p}}{g(\textbf{p})}\right),
\end{equation}
where $g(\cdot)$ denotes the geometric mean of the input. We demonstrate that $\textbf{s}^\star$ differs from $\textbf{s}$ by a constant bias. First, we have
\begin{equation}
\textbf{p} = \text{softmax}(\textbf{s}) = \frac{e^\textbf{s}}{\sum_{i=1}^{|V|}e^{s_i}}
\end{equation}
and
\begin{equation}
g(\textbf{p}) = (\prod_{i=1}^{|V|}p_i)^{1/|V|}.
\end{equation}
Then, we can find that
\begin{equation}
\begin{split}
s_i^\star & = \log\left(\frac{p_i}{g(\textbf{p})}\right) = \log(p_i) - \log(g(\textbf{p}))\\
& = \log(p_i) - \frac{1}{|V|}\sum_{j=1}^{|V|}\log(p_j)\\
& = \log\left(\frac{e^{s_i}}{\sum_{j=1}^{|V|}e^{s_j}}\right) - \frac{1}{|V|}\sum_{j=1}^{|V|}\log\left(\frac{e^{s_j}}{\sum_{k=1}^{|V|}e^{s_k}}\right)\\
& = \left[s_i - \log\sum_{j=1}^{|V|}e^{s_j}\right] - \frac{1}{|V|}\sum_{j=1}^{|V|}\left[s_j-\log\sum_{k=1}^{|V|}e^{s_k}\right]\\
& = s_i - \frac{1}{|V|}\sum_{j=1}^{|V|}s_j,
\end{split}
\end{equation}
which means that each component of $\textbf{s}^\star$ only differs from that of $\textbf{s}$ by the mean of $\textbf{s}$. Therefore, $\textbf{s}^\star - \textbf{s}$ forms a constant vector (bias). 

\subsubsection{Top-$k$ Probabilities}
In case that the API provides the top-$k$ probabilities, which are the $k$ largest elements of $\textbf{p}$, we assume that the provider allows the user to alter token probabilities using a bias through the API. This is reasonable and common in applications, e.g., OpenAI includes this function in their API\footnote{https://help.openai.com/en/articles/5247780-using-logit-bias-to-alter-token-probability-with-the-openai-api}. The bias is added to the specific logits of the tokens before the softmax operation, the API returns the top-$k$ probabilities. Without the loss of generalization, assuming that ${i_1, i_2, \dots, i_m}$ are the indices of selected tokens and $b$ is the bias, the biased probabilities distribution $\textbf{p}^\star$ can be calculated by
\begin{equation}
\textbf{p}^\star(i_1, i_2, \ldots, i_m, b) = \text{softmax}\left(\textbf{s}^\star(i_1, i_2, \ldots, i_m, b)\right),~
\end{equation}
where for $1\leq i\leq |V|$, we have
\begin{equation}
s_i^\star(i_1, i_2, \ldots, i_m, b) =
\begin{cases}
	s_i+b & i \in\{i_1, i_2, \ldots, i_m\}, \\
	s_i & \text{otherwise}.
\end{cases}
\end{equation}
We propose a method to recover the complete probabilities $\textbf{p} = \text{softmax}(\textbf{s})$, i.e., $b \equiv 0$ for Eq. (5). In detail, by feeding a prompt to the model, we can obtain the top-$k$ probabilities, which can be expressed as $\{p_{\text{idx}_1}, p_{\text{idx}_2}, \ldots, p_{\text{idx}_k}\}$, where $p_{\text{idx}_1}\geq p_{\text{idx}_2} \geq \ldots \geq p_{\text{idx}_k}$ and $\{\text{idx}_1, \text{idx}_2, \dots, \text{idx}_k\}$ are the indices of the corresponding tokens. Obviously, we have
\begin{equation}
p_{\text{idx}_i} = \frac{e^{s_{\text{idx}_i}}}{\sum_{j=1}^{|V|}e^{s_j}},~\forall~1\leq i\leq k. 
\end{equation}
We have collected $k$ original probabilities $\{p_{\text{idx}_1}, p_{\text{idx}_2}, \ldots, p_{\text{idx}_k}\}$. We are to determine the remaining $|V|-k$ original probabilities. This problem can be solved by processing these $|V|-k$ tokens in batches each containing $k-1$ tokens. Suppose that we are now to process $k-1$ tokens, whose indices are $\{i_1, i_2, ..., i_{k-1}\}$, we add a bias $b$ to the logits of these $k-1$ tokens and the $\text{idx}_1$-th token to push them into the `top-$k$'. The biased probabilities distribution can therefore be calculated by
\begin{equation}
	\textbf{p}^\star(\text{idx}_1, i_1, \ldots, i_{k-1}, b) = \text{softmax}\left(\textbf{s}^\star(\text{idx}_1, i_1, \ldots, i_{k-1}, b)\right),~
\end{equation}
where for $1\leq i\leq |V|$, we have
\begin{equation}
	s_i^\star(\text{idx}_1, i_1, \ldots, i_{k-1}, b) =
	\begin{cases}
		s_i+b & i \in\{\text{idx}_1, i_1, \ldots, i_{k-1}\}, \\
		s_i & \text{otherwise}.
	\end{cases}
\end{equation}
We can write
\begin{equation}
p_{i_r} = \text{softmax}(s_{i_r}) = \frac{e^{s_{i_r}}}{\sum_{j=1}^{|V|}e^{s_j}},~\forall~i_r \in \{\text{idx}_1, i_1, \ldots, i_{k-1}\},
\end{equation}
and
\begin{equation}
\begin{split}
p_{i_r}^\star & = \text{softmax}(s_{i_r}^\star) = \frac{e^{s_{i_r}^\star}}{\sum_{j=1}^{|V|}e^{s_j^\star}}\\
& = \frac{e^{s_{i_r}^\star}}{
	\sum_{j\notin \{\text{idx}_1, i_1, \ldots, i_{k-1}\}}e^{s_j^\star} + 
	\sum_{j\in \{\text{idx}_1, i_1, \ldots, i_{k-1}\}}e^{s_j^\star}}\\
& = \frac{e^{s_{i_r}+b}}{
	\sum_{j\notin \{\text{idx}_1, i_1, \ldots, i_{k-1}\}}e^{s_j} + 
	\sum_{j\in \{\text{idx}_1, i_1, \ldots, i_{k-1}\}}e^{s_j+b}},
\end{split}
\end{equation}
for any $i_r \in \{\text{idx}_1, i_1, \ldots, i_{k-1}\}$. Obviously, 
\begin{equation}
\frac{p_{\text{idx}_1}}{p_{i_r}} = \frac{e^{s_{\text{idx}_1}}}{e^{s_{i_r}}}
\end{equation}
and
\begin{equation}
\frac{p_{\text{idx}_1}^\star}{p_{i_r}^\star} = \frac{e^{s_{\text{idx}_1}^\star}}{e^{s_{i_r}^\star}} = \frac{e^{s_{\text{idx}_1}+b}}{e^{s_{i_r}+b}} = \frac{e^{s_{\text{idx}_1}}}{e^{s_{i_r}}}.
\end{equation}
Therefore, for any $i_r \in \{i_1, \ldots, i_{k-1}\}$,
\begin{equation}
p_{i_r} = p_{\text{idx}_1}\cdot\frac{p_{i_r}^\star}{p_{\text{idx}_1}^\star},
\end{equation}
indicating that the original probability for the $i_r$-th token $p_{i_r}$ can be recovered since $p_{\text{idx}_1}$, $p_{i_r}^\star$ and $p_{\text{idx}_1}^\star$ are all known. By querying the model with the same prompt, the complete probabilities $\textbf{p}$ can be perfectly reconstructed, according to the above operation.

It is remarked that the above operation processes $k-1$ tokens at each time. In fact, it is free for us to process less then $k$ tokens at each time. Therefore, we should perform at least $(|V| - k) / (k - 1)$ times. In case that $k - 1$ cannot divide $|V| - k$, we need to perform once more. In addition, at each time, we have to choose a suitable value for $b$, which can be done by a heuristic fashion.

\subsubsection{Top-$1$ Probability}
The API returns the highest probability, which is equivalent to obtaining the top-$k$ probabilities with $k = 1$. For this scenario, we present a method to recover the complete probabilities $\textbf{p}$. Let us introduce a substantial bias $b$ to the $i$-th token, thereby elevating it to the top position, resulting in the biased probability $p_i^\star$. The unbiased probability $p_i$ for the $i$-th token can be determined by
\begin{equation}
p_i = \frac{1}{e^{b - \log{p_i^\star}} - e^b + 1}.
\end{equation}
This can be easily verified. First of all, we have
\begin{equation}
p_i = \frac{e^{s_i}}{\sum_{j=1}^{|V|}e^{s_j}}
\end{equation}
and
\begin{equation}
p_i^\star = \frac{e^{s_i+b}}{\sum_{j=1,j\neq i}^{|V|}e^{s_j} + e^{s_i+b}},
\end{equation}
which can be rewritten as
\begin{equation}
\begin{split}
p_i^\star & = \frac{e^{s_i+b}}{\sum_{j=1}^{|V|}e^{s_j} - e^{s_i} + e^{s_i+b}}\\
& = \frac{e^{s_i}\cdot e^b}{\sum_{j=1}^{|V|}e^{s_j} - e^{s_i} + e^{s_i}\cdot e^b}\\
& = \frac{\left(p_i\cdot\sum_{j=1}^{|V|}e^{s_j}\right)\cdot e^b}{\sum_{j=1}^{|V|}e^{s_j} - \left(p_i\cdot\sum_{j=1}^{|V|}e^{s_j}\right) + \left(p_i\cdot\sum_{j=1}^{|V|}e^{s_j}\right)\cdot e^b}\\
& = \frac{p_i\cdot e^b}{1-p_i+p_i\cdot e^b}\\
& = \frac{e^b}{p_i^{-1}-1+e^b}.
\end{split}
\end{equation}
Therefore, we have
\begin{equation}
p_i^{-1} = \frac{e^b}{p_i^\star} - e^b + 1.
\end{equation}

Theoretically, this method could recover the unbiased probability in the top-$k$ scenario, though it would require a substantially larger number of queries. However, its practical applicability is limited due to exponential operations, which introduce numerical instability and can adversely affect subsequent results. In contrast, the method described previously relies solely on proportional operations, thereby largely avoiding numerical instability issues.

\section{Fingerprint Verification}
\begin{figure}[!t]
	\centering
	\includegraphics[width=\linewidth]{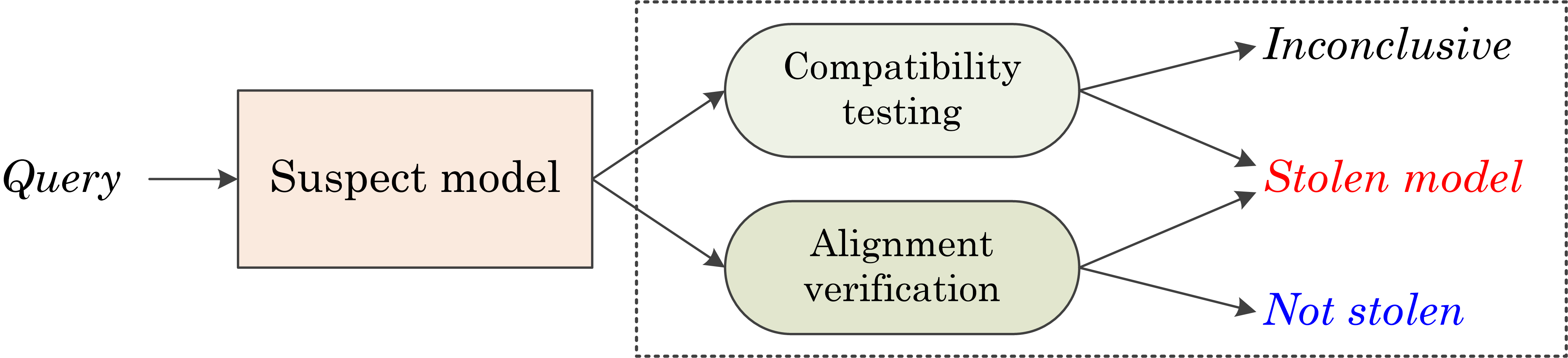}
	\caption{The sketch of LLM fingerprint verification.}\label{sketch}
\end{figure}

We introduce two methods for LLM fingerprint verification. The first method involves verifying whether the outputs of the suspect LLM occupy the same space as those of the victim LLM. It means that the suspect model and the victim model share the same last linear layer, facilitating the rapid identification of model infringement. If the model was fine-tuned, indicating that the parameters of the last linear layer have been modified, i.e., changes occur in the vector space, which makes verification challenging. To deal with this problem, we propose an alignment verification method to resolve this challenge. This method calculates the joint space dimension formed by the output vector space of the suspect model and the parameter space of the victim model. Model infringement is identified by comparing the similarity of these two spaces. Figure \ref{sketch} shows the sketch of LLM fingerprint verification.

\subsection{Compatibility Testing}
As demonstrated above, the LLM outputs span a vector space $L$ isomorphic to the space spanned by the columns of the last linear layer $\textbf{W}$. We slightly abuse the notation by using $L$ to represent the space by the columns of the last linear layer $\textbf{W}$ in the following. We retain the last linear layer of LLM as private, serving as the fingerprint of the victim model. The suspect model is queried through the API to obtain its logits output $\textbf{s}$.

For the suspect model derived from the victim model without modification of the last linear layer, the logits $\mathbf{s}$ is expected to lie within $L$. This can be formally verified by attempting to solve the linear system $\mathbf{W}\mathbf{x} = \mathbf{s}$, to determine whether $\mathbf{s}$ belongs to $L$. If a solution exists, the suspect model is considered to be derived from the victim model. In practice, due to numerical errors introduced by floating-point computations, we instead compute the Euclidean distance $d$ between $\mathbf{s}$ and the subspace $L$ by
\begin{equation}
d = \left\|\mathbf{s} - \mathbf{W}\hat{\mathbf{x}}\right\|,
\end{equation}
where $\hat{\mathbf{x}}$ denotes the solution of $\mathbf{W}\mathbf{x} = \mathbf{s}$. Let $e$ represent the error tolerance induced by numerical instability. If $d < e$, we conclude that $\mathbf{s}$ is compatible with $L$, indicating that the suspect model is likely derived from the victim model.

When only the output probabilities $\mathbf{p}$ are accessible, and logits $\mathbf{s}^\star$ are reconstructed using the method described previously, we account for the inherent one-dimensional deviation by appending a constant column vector to $\mathbf{W}$, yielding $\mathbf{W}^\star = [\mathbf{W},\mathbf{1}]$. The verification procedure is thereafter carried out using $\mathbf{W}^\star$ in place of $\mathbf{W}$, following the same methodology outlined above.

\subsection{Alignment Verification}
Since PEFT alters parts of the model parameters, the verification method described above can no longer be applied reliably. In line with the manifold hypothesis, high-dimensional data can be mapped onto a low-dimensional latent manifold, within which different tasks are controlled by distinct feature directions. Therefore, when fine-tuning for a specific task, it is assumed to impact only a portion of the vector space $L$, rather than the entire space. Especially, in PEFT attacks, the model is fine-tuned with a low-rank matrix. The original matrix $\textbf{W}_\text{O}$ is updated by 
\begin{equation}
\textbf{W}_\text{N} = \textbf{W}_\text{O} + \Delta \textbf{W},
\end{equation}
where $\Delta \textbf{W}$ is a matrix of rank no more than $k$. If there is substantial overlap between between of the output space of the suspect model and the space spanned by the columns of the victim model $\textbf{W}$, the suspect model closely resembles the victim model and may have been derived through unauthorized replication.

We calculate the dimension formed by the union of the vector space of the suspect model and the parameter space of the victim model, denoted by $L_\text{sum}$. If the dimension of $L_\text{sum}$ is quite close to that of the space of the victim model, it indicates that the suspect model is derived from the victim model. Otherwise, it is not. It can be controlled by using a threshold. For those matrices containing numerous floating-point numbers, directly calculating their rank is not advisable, as it leads to significant errors due to numerical inaccuracies. Instead, we
here propose to calculate the dimension difference $\Delta r$ between $L_\text{sum}$ and $L$ (corresponding to $\textbf{W}$) to determine whether the suspect model is derived from the victim model. If $\Delta r$ is smaller than a small threshold (or say $\Delta r$ is significantly smaller than the hidden size $h$), it means strong alignment and supports the inference that the suspect model originates from the victim model. In scenarios where only output probabilities are accessible, $\Delta r$ will only experience a numerical disturbance of one, which will not affect our results. Algorithm 1 gives the pseudocode for dimension difference calculation. 

\begin{algorithm}[!t]
	\caption{Pseudocode for dimension difference calculation}
	\begin{algorithmic}[1]
		\renewcommand{\algorithmicrequire}{\textbf{Input:}}
		\renewcommand{\algorithmicensure}{\textbf{Output:}}
		\REQUIRE The parameter matrix $\textbf{W}$, the logits set $S = \{\textbf{s}_1, \textbf{s}_2, ..., \textbf{s}_q\}$, the error term $e_r$. 
		\ENSURE  The dimension difference $\Delta r$.
		\STATE Initiate $\Delta r = 0$ and $\textbf{W}_\text{sum} = \textbf{W}$
		\FOR {$i$ = 1, 2, ..., $q$}
		\STATE Solve $\textbf{W}_\text{sum}\cdot\textbf{x}_i = \textbf{s}_i$ to obtain $\hat{\textbf{x}}_i$
		\STATE Calculate $d_i = ||\textbf{s}_i - \textbf{W}_\text{sum}\cdot\hat{\textbf{x}}_i||$
		\IF { $d_i > e_r$ }
			\STATE $\Delta r = \Delta r + 1$
			\STATE $\textbf{W}_\text{sum} = \left[\textbf{W}_\text{sum}, \textbf{s}_i\right]$
		\ENDIF
		\ENDFOR
		\RETURN $\Delta r$
	\end{algorithmic}
\end{algorithm}

\section{Experimental Results and Analysis}
In this section, we are to report experimental results and provide analysis to demonstrate the applicability of our work.

\subsection{Setup}
We use Gemma \cite{Mesnard:arXiv:2024} with a hidden size of 2048 as the victim model and fine-tune it by either LoRA \cite{Hu:ICLR:2022} or QLoRA \cite{Dettmers:NIPS:2023} as the  adversarial setting. We set the rank to 16, 32 and 64. The models are fine-tuned on the Alpaca dataset \cite{Dubois:NIPS:2023}  and SAMSum dataset \cite{Gliwa:Workshop:2019} to simulate different scenarios. We also compare with a new version termed as Gemma-2, which shares the same structure but under a novel training method, leading to substantial income and completely different outputs for the identical inputs. For each experiment, we generate
(or recover) 300 complete logits (or probabilities) for simulation. The error parameter $e_r$ is fixed to 1 across all experiments. The following subsections report results for compatibility testing and alignment verification, followed by a case study that illustrates the practical applicability of our method\footnote{Code available: https://github.com/solitude-alive/llm-fingerprint}.

\begin{table*}[!t]
	\caption{~~~~~~~~Table 1. Compatibility testing results on different model versions and models fine-tuned with different modules and ranks.}
	\begin{center}
		\begin{tabular}{c|c|cccc}
			\hline\hline
			\multirow{2}{*}{Model} & \multirow{2}{*}{Dataset} & \multicolumn{4}{c}{Scenario} \\
			& & Full logits & Full probabilities & Top-5 probabilities & Top-1 probabilities\\
			\hline
			{Gemma} & \multirow{2}{*}{/} & {$8.0 \times 10^{-5}$} & {$3.4 \times 10^{-3}$} & {$3.4 \times 10^{-3}$} & {$8.3 \times 10^{-5}$} \\
			{Gemma-2} &  & {$5.5 \times 10^{5}$} & {$5.5 \times 10^{5}$} & {$5.5 \times 10^{5}$} & {$5.5 \times 10^{5}$} \\
			\hline
			{qkv-16} & \multirow{6}{*}{SAMSum} & {$9.7 \times 10^{-5}$} & {$1.4 \times 10^{-4}$} & {$1.4 \times 10^{-4}$} & {$1.0 \times 10^{-4}$} \\
			{qkv-32} & & {$8.5 \times 10^{-5}$} & {$3.6 \times 10^{-3}$} & {$3.6 \times 10^{-3}$} & {$9.6 \times 10^{-5}$} \\
			{qkv-64} & & {$1.1 \times 10^{-4}$} & {$9.5 \times 10^{-4}$} & {$9.5 \times 10^{-4}$} & {$9.9 \times 10^{-5}$} \\
			{linear-16} & & {$7.3 \times 10^{5}$} & {$7.2 \times 10^{5}$} & {$7.2 \times 10^{5}$} & {$7.2 \times 10^{5}$} \\
			{linear-32} & & {$6.9 \times 10^{5}$} & {$6.8 \times 10^{5}$} & {$6.8 \times 10^{5}$} & {$6.9 \times 10^{5}$} \\
			{linear-64} & & {$7.6 \times 10^{5}$} & {$7.5 \times 10^{5}$} & {$7.5 \times 10^{5}$} & {$7.5 \times 10^{5}$} \\
			\hline
			{qkv-16} & \multirow{6}{*}{Alpaca} & {$8.3 \times 10^{-5}$} & {$5.7 \times 10^{-4}$} & {$5.7 \times 10^{-4}$} & {$8.4 \times 10^{-4}$} \\
			{qkv-32} & & {$7.0 \times 10^{-5}$} & {$2.8 \times 10^{-3}$} & {$2.8 \times 10^{-3}$} & {$8.5 \times 10^{-5}$} \\
			{qkv-64} & & {$6.8 \times 10^{-5}$} & {$1.6 \times 10^{-4}$} & {$1.6 \times 10^{-4}$} & {$8.1 \times 10^{-5}$} \\
			{linear-16} & & {$6.8 \times 10^{5}$} & {$6.7 \times 10^{5}$} & {$6.7 \times 10^{5}$} & {$6.7 \times 10^{5}$} \\
			{linear-32} & & {$6.7 \times 10^{5}$} & {$6.6 \times 10^{5}$} & {$6.6 \times 10^{5}$} & {$6.6 \times 10^{5}$} \\
			{linear-64} & & {$6.9 \times 10^{5}$} & {$6.8 \times 10^{5}$} & {$6.8 \times 10^{5}$} & {$6.9 \times 10^{5}$} \\
			\hline\hline
		\end{tabular}
	\end{center}
\end{table*}

\begin{table*}[!t]
	\caption{~~~~~~~~Table 2. Dimension difference results on different model versions and models fine-tuned with different modules and ranks.}
	\begin{center}
		\begin{tabular}{c|c|cccc}
			\hline\hline
			\multirow{2}{*}{Model} & \multirow{2}{*}{Dataset} & \multicolumn{4}{c}{Scenario} \\
			& & Full logits & Full probabilities & Top-5 probabilities & Top-1 probabilities\\
			\hline
			{Gemma} & \multirow{2}{*}{/} & {0} & {1} & {1} & {1} \\
			{Gemma-2} &  & {300} & {300} & {300} & {300} \\
			\hline
			{qkv-16} & \multirow{6}{*}{SAMSum} & {0} & {1} & {1} & {1} \\
			{qkv-32} & & {0} & {1} & {1} & {1} \\
			{qkv-64} & & {0} & {1} & {1} & {1} \\
			{linear-16} & & {10} & {11} & {11} & {11} \\
			{linear-32} & & {11} & {12} & {12} & {11} \\
			{linear-64} & & {10} & {11} & {11} & {11} \\
			\hline
			{qkv-16} & \multirow{6}{*}{Alpaca} & {0} & {1} & {1} & {1} \\
			{qkv-32} & & {0} & {1} & {1} & {1} \\
			{qkv-64} & & {0} & {1} & {1} & {1} \\
			{linear-16} & & {15} & {16} & {16} & {16} \\
			{linear-32} & & {20} & {21} & {21} & {20} \\
			{linear-64} & & {21} & {22} & {22} & {21} \\
			\hline\hline
		\end{tabular}
	\end{center}
\end{table*}

\begin{table*}[!t]
	\caption{~~~~~~~~~Table 3. Dimension difference results for models fine-tuned with QLoRA at different ranks across different datasets.}
	\begin{center}
		\begin{tabular}{c|c|cccc}
			\hline\hline
			\multirow{2}{*}{Model} & \multirow{2}{*}{Dataset} & \multicolumn{4}{c}{Scenario} \\
			& & Full logits & Full probabilities & Top-5 probabilities & Top-1 probabilities\\
			\hline
			{Gemma} & \multirow{2}{*}{/} & {0} & {1} & {1} & {1} \\
			{Gemma-2} &  & {300} & {300} & {300} & {300} \\
			\hline
			{qkv-16} & \multirow{6}{*}{SAMSum} & {0} & {1} & {1} & {1} \\
			{qkv-32} & & {0} & {1} & {1} & {1} \\
			{qkv-64} & & {0} & {1} & {1} & {1} \\
			{linear-16} & & {9} & {10} & {10} & {10} \\
			{linear-32} & & {10} & {11} & {11} & {11} \\
			{linear-64} & & {8} & {9} & {9} & {9} \\
			\hline
			{qkv-16} & \multirow{6}{*}{Alpaca} & {0} & {1} & {1} & {1} \\
			{qkv-32} & & {0} & {1} & {1} & {1} \\
			{qkv-64} & & {0} & {1} & {1} & {1} \\
			{linear-16} & & {16} & {17} & {17} & {17} \\
			{linear-32} & & {20} & {21} & {21} & {21} \\
			{linear-64} & & {23} & {24} & {24} & {24} \\
			\hline\hline
		\end{tabular}
	\end{center}
\end{table*}

\subsection{Compatibility Testing Results}
In the compatibility testing, we assume that malicious users might release their stolen model either in its original form or after fine-tuning. Fine-tuning is applied to the attention mechanism in the intermediate layers or the last linear layer. Here, we retain the parameters of the model's last linear layer and conduct verification experiments across various scenarios. The average Euclidean distance $d$ is used to quantify the performance. As shown in Table 1, each column corresponds to the results of $d$ under different API scenarios, while each row represents the outcomes for different models. `qkv' and `linear' represent LoRA module was applied to the query, key and value module in the attention mechanism or the linear module in the last layer, with their suffixes indicating the rank of LoRA. The experimental results clearly reveal that for models fine-tuned on the last layer or not, a significant difference exits between them and other models. This distinction is adequate for ascertaining ownership, verifying the effectiveness of the proposed method. It is worth to note that although this is the average results, this consistency will hold even with only a few samples in practice, thus also enabling rapid verification.

\subsection{Alignment Verification Results}
In case of alignment verification, Table 2 presents the results of the dimension difference calculation, i.e., $\Delta r$. The configuration of PEFT attacks is the same as that in the previous subsection. As shown in Table 2, the proposed method consistently distinguishes between different types of fine-tuning across all four API scenarios. For models fine-tuned on attention modules, the metric returns uniformly small $\Delta r$,  indicating strong consistency with the original model. In contrast, models fine-tuned on the final linear layer exhibit substantially larger $\Delta r$ that grows systematically with LoRA rank. All these observations are not merely highlighting the difference between qkv and linear fine-tuning; rather, they directly demonstrate the discriminative power of our method. Regardless of the fine-tuning location, fine-tuning intensity, or the granularity of API outputs, our method reliably captures the structural deviations introduced by PEFT, enabling effective identification of stolen or fine-tuned variants. The consistent trends observed across all API scenarios further validate the robustness and sample efficiency of the proposed verification method. Table 3 reports the dimension difference results for models fine-tuned with QLoRA at different ranks across different datasets. From this table, we can draw out the same conclusions.

It is noted that $\Delta r = 300$ reported for Gemma-2 in Table 2  and Table 3 is a result of the fact that we utilize only 300 valid complete logits (or probabilities) for simulation. If this restriction is relaxed and more outputs are included, the observed $\Delta r$ would likely increase proportionally. Indeed, analysis of 3000 outputs from Gemma-2 yields $\Delta r = 2052$, which suggests extensive fine-tuning or even training from scratch. This further emphasizes the uniqueness and reliability of the proposed fingerprint.

\subsection{Case Study}
\begin{figure}[!t]
	\centering
	\includegraphics[width=\linewidth]{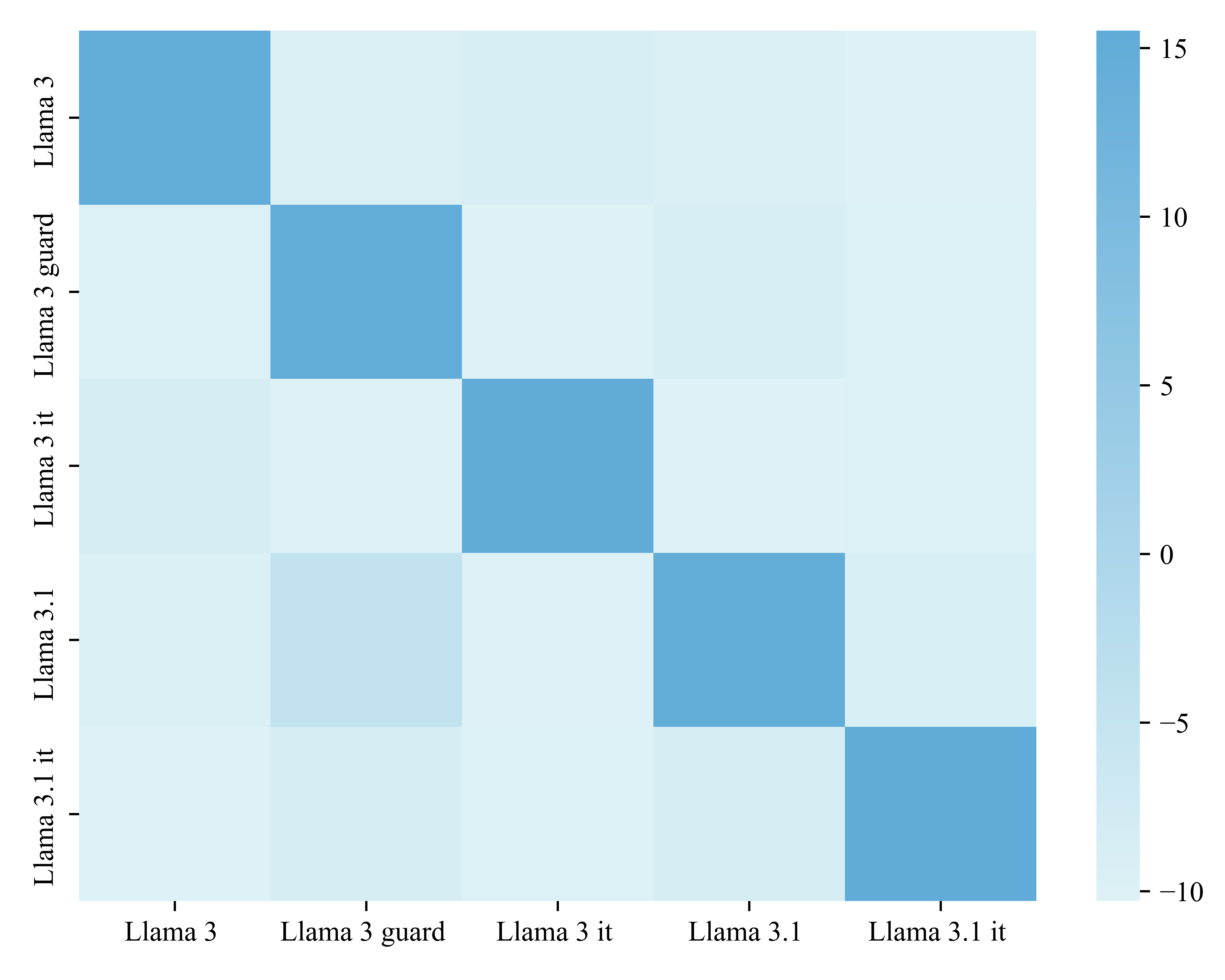}
	\caption{A case study on Llama-family models.}\label{llama}
\end{figure}

We conduct a case study on five Llama-family models sharing same architectures to further evaluate the effectiveness of our fingerprinting method under realistic conditions of high model similarity. In Figure \ref{llama}, the contribution logits and weights, computed via the Euclidean distance metric and visualized after logarithmic transformation, exhibit a clear and stable pattern, that is, only the diagonal entries show markedly elevated values, while all off-diagonal entries remain uniformly low. This indicates that our metric consistently identifies each model as most compatible with itself and effectively rejects all other closely related variants. Crucially, this pronounced diagonal dominance does not arise from subtle architectural differences but directly reflects the discriminative strength and reliability of our method. Even among models with identical architectures and highly similar training objectives, the resulting fingerprints remain distinctive and stable, enabling robust and trustworthy model attribution in practical applications.

\section{Conclusion and Discussion}
In this paper, we introduce a novel and general fingerprinting framework for large language models that enables reliable ownership verification in black-box settings without retraining or modifying the target model. Across multiple model families, API granularities, and PEFT attack configurations, our experiments show that the proposed method produces distinctive and stable fingerprints. The approach consistently achieves high verification accuracy and remains robust to LoRA fine-tuning on various modules. These findings highlight persistent structural properties of LLMs that enable dependable model attribution even under adversarial conditions. Overall, this research offers a practical and effective solution for LLM provenance tracking and ownership protection, and outlines a promising direction for future research on secure deployment of generative models.

\section{Acknowledgment} 
This study was financially supported by the Nanning ``Yong Jiang'' Program under Grant Number RC20250102, Science and Technology Commission of Shanghai Municipality under Grant Number 24ZR1424000, and Xizang Autonomous Region Central Guided Local Science and Technology Development Fund Project under Grant Number XZ202401YD0015. 




\begin{biography}
Zhiguang Yang received his BS and MS degrees from Shanghai University, Shanghai, China, in 2022 and 2025, respectively. His research interests include deep learning, large language models, digital watermarking and fingerprinting. He is now an Algorithm Engineer at a startup.

Hanzhou Wu received his BS and PhD degrees from Southwest Jiaotong University, Chengdu, China, in 2011 and 2017, respectively. He was a Visiting Scholar in New Jersey Institute of Technology, New Jersey, USA, from 2014 to 2016. He was a Research Scientist in Institute of Automation, Chinese Academy of Sciences, Beijing, China, from 2017 to 2019. He is now an Associate Professor in Shanghai University, Shanghai, China. His research interests include steganography, steganalysis, digital watermarking and digital forensics. He has published more than 100 research articles in peer journals and conferences. He has also written four book chapters. He served as the Organization Chair for 2022 IEEE International Workshop on Information Forensics and Security, and serves as an Associate Editor for IEEE Signal Processing Letters started from 2025. 
\end{biography}

\end{document}